\documentclass[final, aps, pra, floatfix, reprint, nofootinbib, citeautoscript, superscriptaddress, nobalancelastpage]{revtex4-2}
\usepackage[T1]{fontenc}
\usepackage[english]{babel}

\usepackage{amsmath}
\usepackage{amssymb}
\usepackage{bm}
\usepackage{lmodern}
\usepackage{textcomp}

\usepackage[normalem]{ulem}

\usepackage[colorlinks, urlcolor=blue, citecolor=blue, linkcolor=blue, pdfstartview=FitH]{hyperref}
\usepackage[all]{hypcap}
\usepackage[dvipsnames]{xcolor}
\usepackage{subfigure}
\usepackage{graphicx}

\def\affiSOLAB{Spin\ Optics\ Laboratory, Saint~Petersburg\ State\ University, 198504 St.~Peterbsurg, Russia}
\def\affiIOFFE{Ioffe\ Institute, 194021 St.~Petersburg, Russia}

\begin{document}
\author{A.\ A.\ Fomin}
\affiliation{\affiSOLAB}

\author{M.\ Yu.\ Petrov}
\affiliation{\affiSOLAB}

\author{G.\ G.\ Kozlov}
\affiliation{\affiSOLAB}

\author{A.\ K.\ Vershovskii}
\affiliation{\affiIOFFE}

\author{M.\ M.\ Glazov}
\affiliation{\affiIOFFE}

\author{V.\ S.\ Zapasskii}
\affiliation{\affiSOLAB}

\title{Anomalous light-induced broadening of the spin-noise resonance in cesium vapor}
\begin{abstract}
We uncover a highly nontrivial dependence of the spin-noise (SN) resonance broadening induced by the intense probe beam. The measurements were performed by probing the cell with cesium vapor at the wavelengths of the transition ${6}^2S_{1/2} \leftrightarrow {6}^2P_{3/2}$ ($\mathrm{D}_2$ line) with the unresolved hyperfine structure of the excited state. 
The light-induced broadening of the SN resonance was found to differ strongly at different slopes of the $\mathrm{D}_2$ line and, generally, varied nonmonotonically with light power. 
We discuss the effect in terms of the phenomenological Bloch equations for the spin fluctuations and demonstrate that the SN broadening behavior strongly depends on the relation between the pumping and excited-level decay rates, the spin precession, and decoherence rates. 
To reconcile the puzzling experimental results, we propose that the degree of optical perturbation of the spin-system is controlled by the route of the excited-state relaxation of the atom or, in other words, that the act of optical excitation of the atom does not necessarily break down completely its ground-state coherence and continuity of the spin precession. 
Spectral asymmetry of the effect, in this case, is provided by the position of the ``closed'' transition $F = 4 \leftrightarrow F' = 5$  at the short-wavelength side of the line. 
This hypothesis, however, remains to be proven by microscopic calculations.
\end{abstract}

\maketitle

\section{Introduction}\label{sec:Introduction}

Spin noise spectroscopy (SNS) is a developing method of magnetic resonance research that is aimed to detect spontaneous (rather than induced) precession of electron spins. 
Spontaneous or incoherent precession of the spin system in its equilibrium state is revealed in the form of fluctuating magnetization which may be detected as fluctuations of magneto-optical effects (Faraday or Kerr rotation). 
The viability of this effect was first demonstrated in 1981~\cite{aleksandrov81}, by detecting magnetic resonance in the Faraday-rotation noise spectrum of sodium atoms. 
The mainstream of research in the field of SNS has arisen in our century after successful detection of spin-noise resonances in semiconductors~\cite{Oestreich_noise, PhysRevB.79.035208}. 
Nowadays, the SNS is widely used not only for studying magnetic resonance and spin dynamics of diverse paramagnets, as a specific method of electron paramagnetic resonance (EPR) spectroscopy, but also for studying properties of optical transitions~\cite{PhysRevLett.110.176601, Yang:2014aa}, spatial spin distribution\ \cite{doi:10.1063/1.3098074, PhysRevB.81.121202}, nuclear-spin dynamics in semiconductor structures~\cite{doi:10.1063/1.4922771, PhysRevB.95.125312, 2015arXiv150605370B}, spin alignment noise in atomic systems~\cite{PhysRevResearch.2.012008, Kozlov:21}, etc., see Refs.~\cite{Oestreich:rev, Zapasskii:13, 2016arXiv160306858S, glazov2018electron, smirnov:SNS:rev} for a review. 
\looseness=1

An important feature of the SNS, which was primarily considered as the most valuable property of this technique, is that the probe laser beam, providing information about spin-system fluctuations, propagates through the medium in the spectral region of its transparency, does not induce any real transitions, and, therefore, does not perturb the spin system. 
In the last decade, however, it was shown that the SNS with resonant or high-power probing, when the probe beam may noticeably affect properties of the material, realizes the \emph{resonant} or \emph{nonlinear} version of this technique with much broader capabilities.
Under these conditions, the SNS becomes essentially perturbative, with characteristics of this perturbation being the subject of investigation, see, e.g., Refs.~\cite{Li:2013fk, PhysRevLett.113.156601, glazov:sns:jetp16, PhysRevB.95.241408, PhysRevB.97.081403}.
\looseness=1

While various aspects of perturbations in the SNS of semiconductors and semiconductor nanosystems are widely studied, see, e.g., Refs.~\cite{PhysRevLett.113.156601, glazov:sns:jetp16, Ryzhov:2016aa, PhysRevB.95.241408, PhysRevB.97.081403, PhysRevB.98.125426, Versh:2020}, much less is known about these effects in atomic systems. 
In Refs.~\cite{PhysRevA.84.043851, PhysRevLett.113.156601}, the nonlinearities related to the coherent and collective effects in the spin system in alkali-metal atoms have been studied.
Previous work~\cite{PhysRevA.103.023104} studied the effects of the probe-beam-induced renormalizations of the spin system's energy spectrum in a strong optical field.
Significant prospects are related to the realization of the spin correlations by the perturbative SNS~\cite{Kong:2020aa}.
 
This paper is devoted to studying what may seem to be one of the simplest nonlinear effects of SNS, namely, to the light-induced broadening of the spin-noise (SN) resonance in the field of the resonant probe beam. 
Our interest in this problem was initiated by a curious spectral and intensity-related behavior of this effect in cesium vapor probed in the region of the transition $6^2S_{1/2} \leftrightarrow 6^2P_{3/2}$ ($\mathrm{D}_2$ line). 
The broadening was found to be essentially different at different sides of the transition, varying strongly nonmonotonously with light power at one of them. 

In addition, the effect showed unusual dependence on the density and/or temperature of the atomic system. 

Phenomenological analysis of the experimental data has shown that the observed anomalous behavior of the SN resonance broadening could be unambiguously explained under the assumption that the ground-state spin precession was more efficiently perturbed in the region of the ``open'' optical transitions, when the atom, after the excitation cycle, did not necessarily return to its initial state. 

The paper is organized as follows. Section~\ref{sec:setup} describes experimental setup and Cs vapor sample. 
Section~\ref{sec:puzzles} summarizes all the experimental observations that attracted our attention and Sec.~\ref{sec:disc} discuss these results. 
A brief conclusion is presented in Sec.~\ref{sec:concl}.

\section{Experimental setup and sample}\label{sec:setup}

\begin{figure}[t]  
	\includegraphics[width=\linewidth]{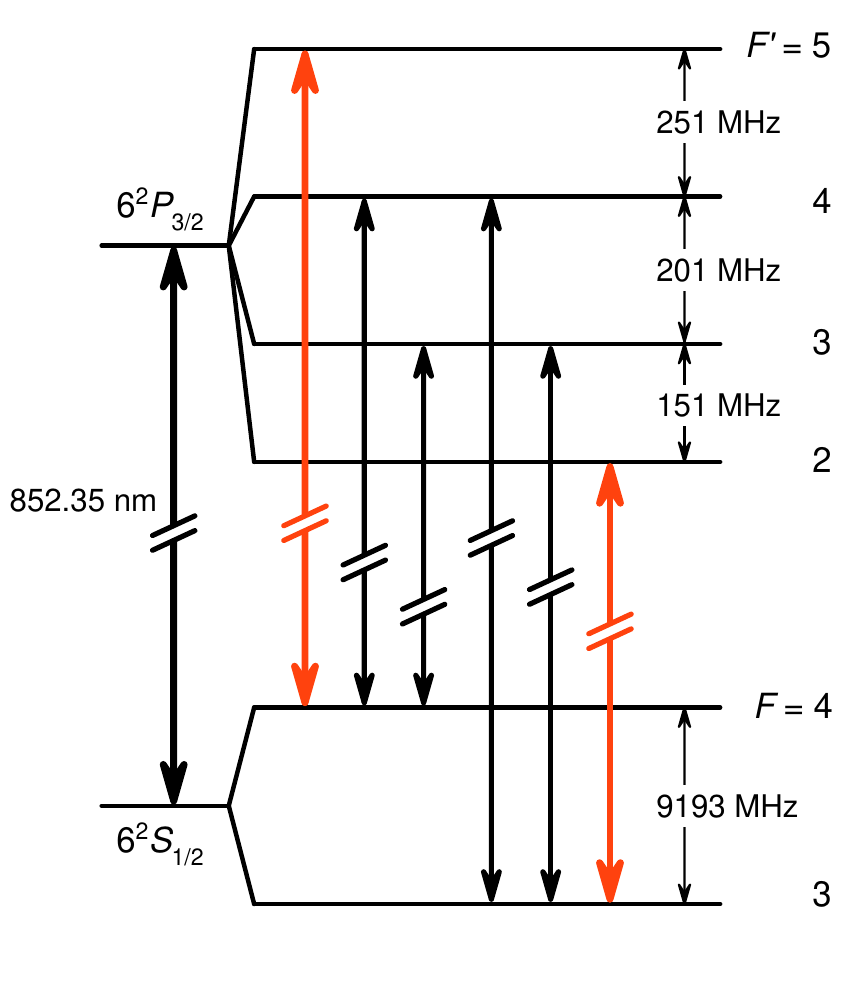}
	\caption{The energy-level diagram of the D$_2$ line of the cesium atom. The ``closed'' transitions are indicated in red.%
	}\label{fig1}
\end{figure}

The measurements were performed at the wavelength of the $\mathrm{D}_2$ line of the cesium atom corresponding to the transitions $6^2S_{1/2}$  ($F = 3, 4$) $\leftrightarrow 6^2P_{3/2}$ ($F' = 2,\ldots, 5$), $\lambda \approx 852.35$~nm. 
The relevant energy-level diagram of the cesium atom is presented in Fig.~\ref{fig1}.

In our experiments, the Doppler width of the $\mathrm{D}_2$ line is comparable with the hyperfine (HF) splitting of the excited state, so that, in the absorption spectrum of this line, one can observe only two spectral components, corresponding to transitions from two HF components of the ground state (Fig.~\ref{fig1})~\cite{PhysRevA.97.032502}. 
In accordance with the selection rules, the allowed transitions from the states $F = 3$ and $F = 4$ may occur to the excited states with the total spin $F' = 2, 3, 4$ and $F' = 3, 4, 5$, respectively.

\begin{figure}  
    \includegraphics[width=\linewidth]{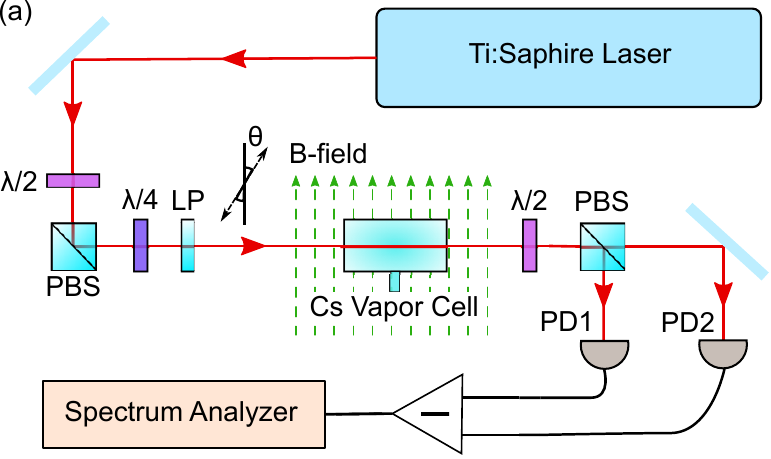}
	\includegraphics[width=\linewidth]{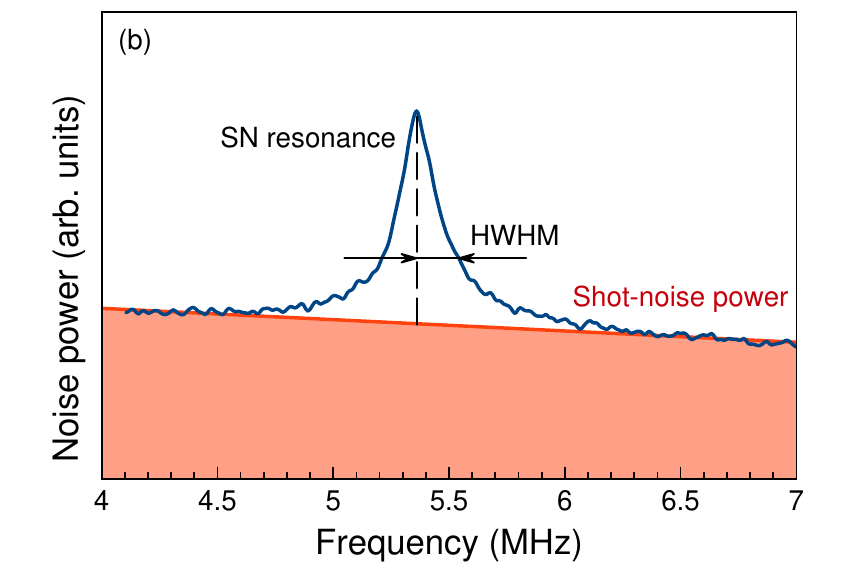}
	\caption{(a) Schematic of the experimental setup. Here, $\lambda/2$ and $\lambda/4$ are the half- and quarter-wave plates, PBS is the polarizing beamsplitter, LP is the linear polarizer, PD1 and PD2 are the photodiodes. The frequency of the strong linearly polarized laser light was scanned in the vicinity of the transitions $F = 4  \leftrightarrow F' = 3, 4, 5$ of the cesium $\mathrm{D}_2$ line.  (b) A typical experimental SN spectrum of cesium under our experimental conditions. A slight tilt of the shot-noise power spectrum is related to the frequency response characteristic of the detection channel.%
	}\label{fig2}
\end{figure}

The SN spectra were detected in the conventional Voigt geometry with the probe beam tuned in resonance with the transition from the upper HF component of the ground state ($F = 4$). 
We also confirmed that similar results were obtained when the lower HF component of the ground state with $F = 3$ was probed. 
A schematic of the experimental setup is shown in Fig.~\ref{fig2}(a). 
As a light source, we used the ring-cavity tunable Ti:sapphire laser with Fabry-Perot-based frequency stabilization. 
This laser allowed us to scan the probe frequency with a step of 100 MHz and thus to study the wavelength dependence of the SN spectra with a spectral resolution substantially exceeding the Doppler width of the $\mathrm{D}_2$ line. 
Fluctuations of the polarization plane azimuth of the transmitted laser beam were detected with a balanced polarimetric detector and processed with a Fourier-transform spectrum analyzer, providing at the output the SN power spectrum of the system. 
The power of the laser beam, $\approx$$4$~mm in diameter, was varied from tens of $\mu$W to several mW. 
To avoid the drift of the laser intensity it was stabilized to $\approx$$0.03\%$ by a liquid-crystal laser power controller (not shown in the figure). 
The cell with a small amount of metal cesium and $2$~torr of buffer gas (Ne), $\varnothing 20 \times 20$~mm in size, was placed into a heater that provided its temperature stabilization at the level of $\pm 1$~\textcelsius{}. 
The magnetic field of around $1.6$~mT, with homogeneity of $\approx$$1$\% within the cell, was created by a pair of Helmholtz-type coils.
\looseness=1

A typical SN spectrum of Cs atoms obtained under these experimental conditions is shown in Fig.~\ref{fig2}(b). 
The SN power, in most cases, exceeded the shot-noise level by more than $10$\%. 
The lineshape was usually approximated by a Lorentzian (see Sec.~\ref{subsec:shape} where the shape is discussed in more detail), with the half-width at half-maximum (HWHM) taken for the measure of its width. 
We stress that, despite elevated probe power densities used in our experiments, the detected noise signal is indeed related to the precession of the Cs atoms spins: The SN power is peaked at the Larmor frequency, see Fig.~\ref{fig2}(b) and also Fig.~\ref{fig7}, and we do not observe any critical behavior being a hallmark of any maser-like effect in our system.

Note that experimental conditions of the present investigation differ substantially from those of our previous work \cite{PhysRevA.103.023104}, which was devoted to studying the light-induced energy splitting of the cesium ground-state spin system.
These two experimental works are indeed close methodologically but, in the previous study~\cite{PhysRevA.103.023104}, the measurements were performed at the long-wavelength slope of the line, where, despite even higher power densities of the probe beam, as will be shown below, the light-induced broadening is almost absent.
It is also noteworthy that the effects of the SN resonance broadening described in this paper did not show any pronounced dependence on the azimuth of the light-beam polarization plane with respect to the magnetic field. 

\begin{figure}[t]
    \includegraphics[width=\linewidth]{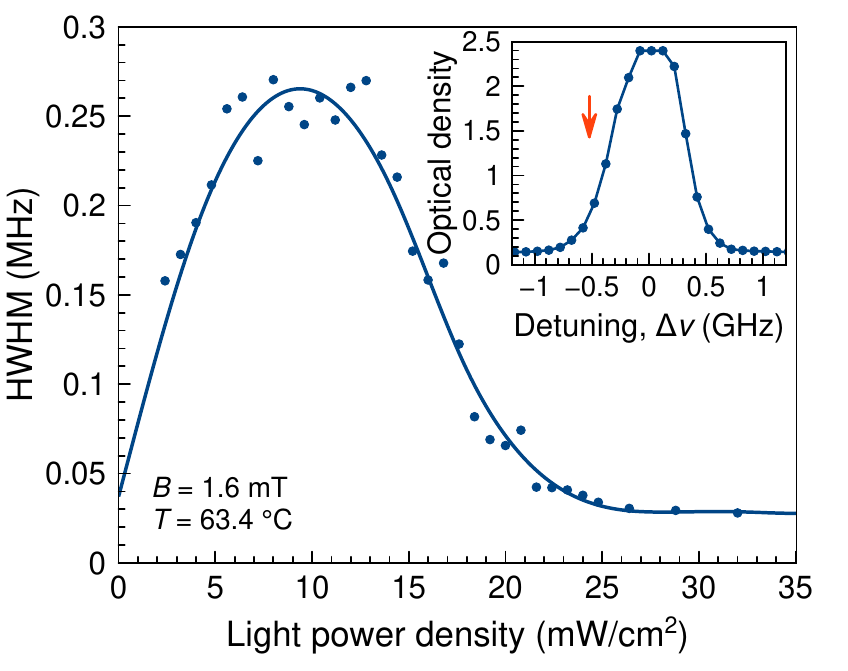}
    \caption{Experimental dependence of the light-induced broadening of the SN resonance in cesium vapor. Here and in the following experimental figures, points show the data, and solid lines are guides for the eye. The spectrum of linear absorption of the $\mathrm D_2$ line is shown in the inset, with the vertical arrow indicating the spectral position of the probe light. The optical spectrum of the SN power, under conditions of linear probing, is shown in Fig.~\ref{fig5}(b). %
	}\label{Fig3}
\end{figure}  

\section{Experimental observations}\label{sec:puzzles} 

In this section, we report on the puzzling variation of the light-induced broadening of the SN resonance as a function of intensity (Sec.~\ref{subsec:int}) and detuning (Sec.~\ref{subsec:det}) of the probe laser beam and versus the cell temperature (Sec.~\ref{subsec:temp}). 
The shape of the SN resonance is discussed in Sec.~\ref{subsec:shape}.

\subsection{Intensity-related variations of the SN resonance width}\label{subsec:int}

The first experimental observation that attracted our attention to this issue was a nonmonotonous dependence of the SN resonance width on light intensity for the probe beam tuned to the long-wavelength slope of the transition $F = 4 \leftrightarrow F' = 3, 4, 5$ (Fig.~\ref{Fig3}). 
The SN resonance profile, in these experiments, was fit with a single Lorentzian, although, more accurately, as will be seen below, it should be considered as comprised of two components, with the width of one of them being intensity-independent. 
At low light-power densities, i.e., in the left part of the plot in Fig.~\ref{Fig3}, the SN linewidth increased monotonically, which seemed quite natural. 
Indeed, the increasing excitation rate strengthened the perturbation of the cesium spin-system, shortened the spin dephasing time in the ground state of the cesium atom, and, thus, broadened its spin-resonance peak. 
However, further nonmonotonic behavior of the SN resonance width looked paradoxical: at sufficiently high light-power densities (of around 20 mW/cm$^2$), the width of the SN resonance became, to within the experimental error, equal to that of the unperturbed spin-system observed either at lowest intensities of the resonant probe beam or at strongly nonresonant probing. 
\looseness=1

\begin{figure}[t]
	\includegraphics[width=\linewidth]{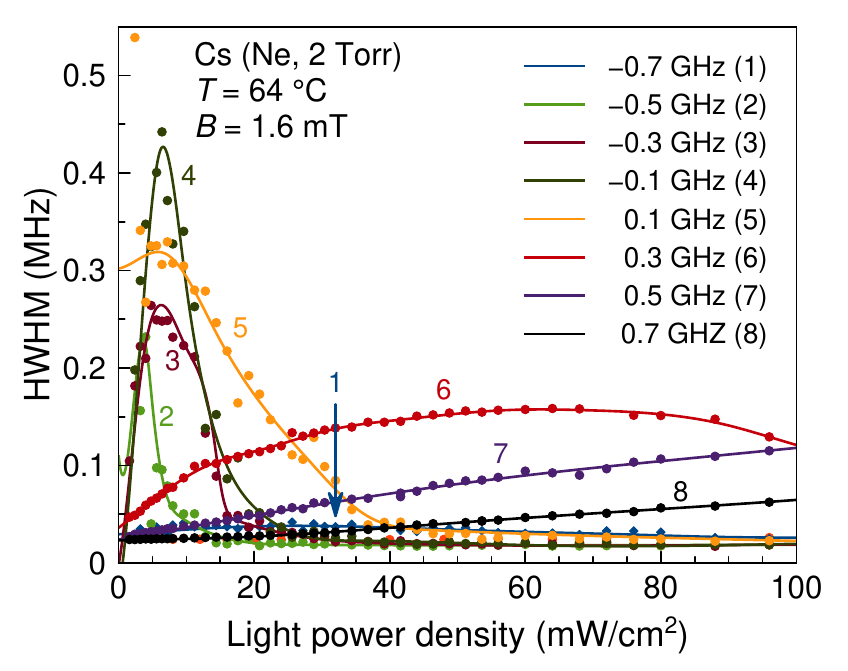}
	\caption{Dependence of the SN resonance width on the probe beam intensity at different detunings of the probe light frequency from the center of the Doppler-broadened transition $F = 4 \leftrightarrow F' = 3, 4, 5$ of cesium.%
	}\label{fig4}
\end{figure}

\subsection{Spectral characteristics of the effect}\label{subsec:det}

To get additional information about this nonmonotonic, anomalous intensity dependence of the SN resonance width, we examined how this dependence varied with the wavelength of the probe light within the $\mathrm{D}_2$ linewidth.
Results of these measurements are shown in Fig.~\ref{fig4}. 
As seen from the figure, the effects of the light-induced SN resonance broadening are strongly different at different sides of the $\mathrm{D}_2$ line. 
The effect of optical perturbations at the short-wavelength (positive detuning) side of the line appeared to be much weaker and practically monotonic with the light intensity, while the anomalous behavior was most pronounced at the long-wavelength side of the resonance (negative detuning).

\begin{figure}[t]
    \includegraphics[width=\linewidth]{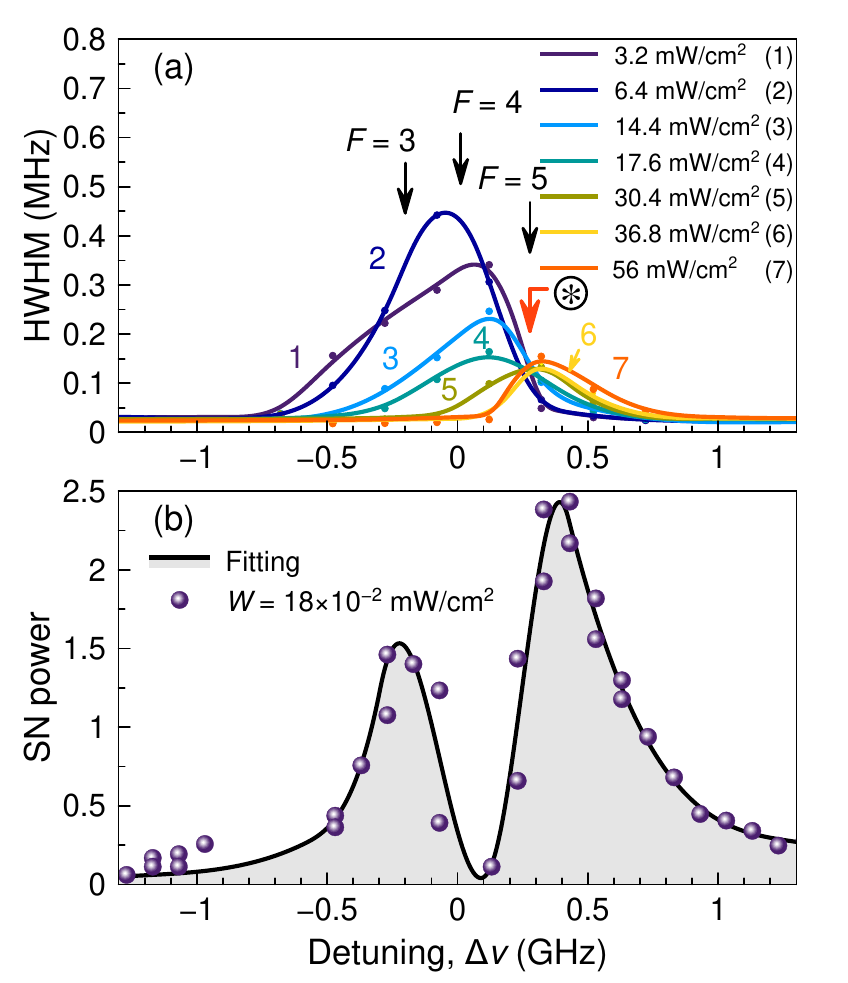}
    \caption{(a) Spectral dependence of the SN resonance width of cesium vapor at different intensities of the probe beam. Arrows above the curves indicate positions of the excited-state HF components. A spectral point where the SN resonance width practically does not depend on the light intensity is indicated by the asterisk. Panel (b) shows, for comparison, the optical spectrum of the SN power of cesium atoms under conditions of weak (linear) probing (taken from our earlier publication~\cite{PhysRevA.97.032502}). This spectrum shows that, in the linear regime, the optical spectrum of the SN does not reveal any hidden HF structure of the transition, see also inset in Fig.~\ref{Fig3}.%
    }\label{fig5}
\end{figure} 

These dependences look even more spectacular as a function of the optical frequency (for fixed probe beam intensities). 
Figure~\ref{fig5} shows, in the same scale along the abscissa axis, optical spectra of the SN resonance broadening at different light intensities (a) and optical spectrum of the SN power of Cs vapor in the regime of linear probing reported in~\cite{PhysRevA.97.032502} [Fig.~\ref{fig5}(b)].
The spectra of the SN resonance broadening, as seen from Fig.~\ref{fig5}(a), are clearly separated into two unequal parts, with the larger (long-wavelength) part being (at low intensities) much more susceptible to optical perturbation. 
Curiously, due to essentially different behavior of the SN resonance broadening at different wings of the line, there exists a point (indicated by the asterisk in Fig.~\ref{fig5}), where the SN resonance width appears to be virtually independent of the probe beam intensity in a wide range of its variation.   

\begin{figure}[t]
    \includegraphics[width=\linewidth]{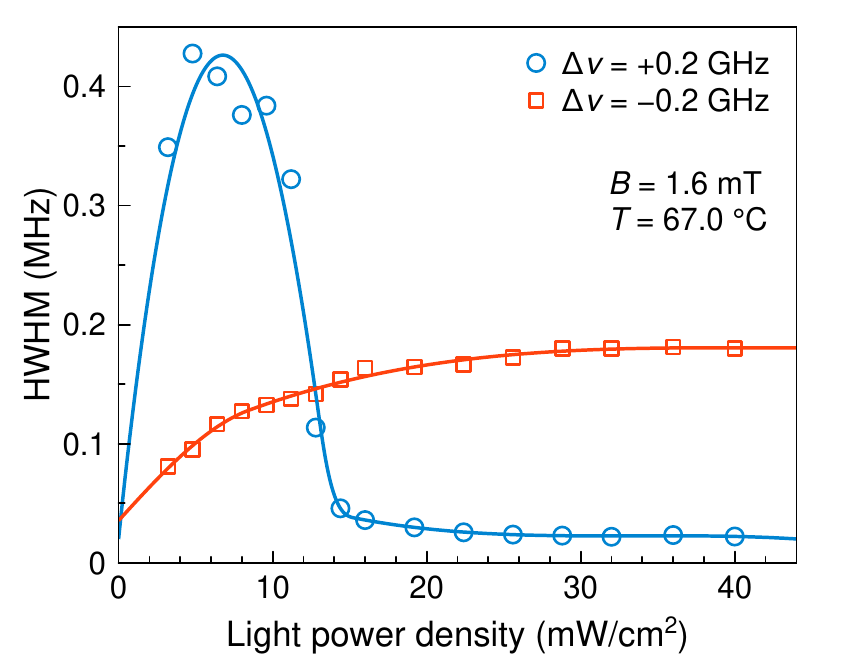}
    \caption{Example of the same dependence as in Fig.~\ref{fig4}, but obtained on the transition from the lower HF component of the ground state ($F=3$). The symbols represent the data measured at $+0.2$~GHz (circles) $-0.2$~GHz (squares) detunings. The solid lines are a guide to the eye. The sides of the transition with higher and lower sensitivity to optical perturbation of the ground-state spin-system are interchanged.%
    }\label{fig6}
\end{figure}

It is important to note that the other HF component of the $\mathrm{D}_2$ line, corresponding to the transition from the lower sublevel ($F = 3$), exhibits qualitatively the same behavior with inversion of the short- and long-wavelength sides of the absorption line, as shown in Fig.~\ref{fig6}. 
In this case, the short-wavelength part is more susceptible to optical perturbation than the long-wavelength one.

\subsection{Shape of the SN resonance}\label{subsec:shape}

\begin{figure}[bt]
    \includegraphics[width=\linewidth]{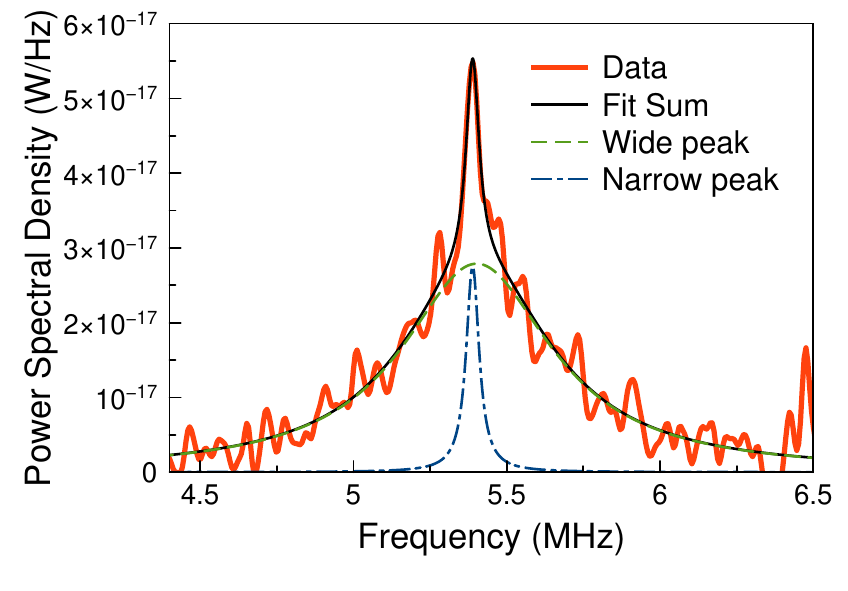}
    \caption{A typical shape of the Faraday rotation noise (SN) spectrum of cesium vapor (red curve) and its decomposition into two Lorentzians: dashed and dash-dotted curves. The fitting sum is shown by the thin black line. Note that the contributions of the electronic and photon shot noise have been subtracted.%
    }\label{fig7}
\end{figure}

In all the measurements presented above, the SN resonance was approximated by a single Lorentzian. 
More accurate analysis has shown, however, that the profile of the  SN resonance,  practically in all cases, can be presented as a sum of two Lorentzians with different widths. A typical example of such a decomposition is shown in Fig.~\ref{fig7}. 
We have found that the width of the narrow component of the SN resonance, within the experimental error, did not depend on the light intensity. Based on this fact, we ascribed this component to SN of the lower HF sublevel of the ground state ($F = 3$). 
Large detuning from the resonant transition provides, in this case, weak perturbation of the fluctuating spins.  
This supposition was confirmed by the measurements of the SN spectra at frequencies detuned by $\approx$$9.2$~GHz above the transition from the component $F = 3$. 
The SN line, under these conditions, could be easily detected, and its width did not noticeably depend on the probe beam power.

In most cases, this contribution (from the lower HF sublevel of the ground state) did not affect essentially results of the measurements, and all the dependences presented above in Figs.~\ref{Fig3}--\ref{fig5} remained qualitatively the same for the ``wide'' component of the SN resonance related, as we believe, to SN of the resonantly probed state with $F = 4$. 

\subsection{Temperature variations of the effect}\label{subsec:temp}

\begin{figure}[tb]
	\includegraphics[width=\linewidth]{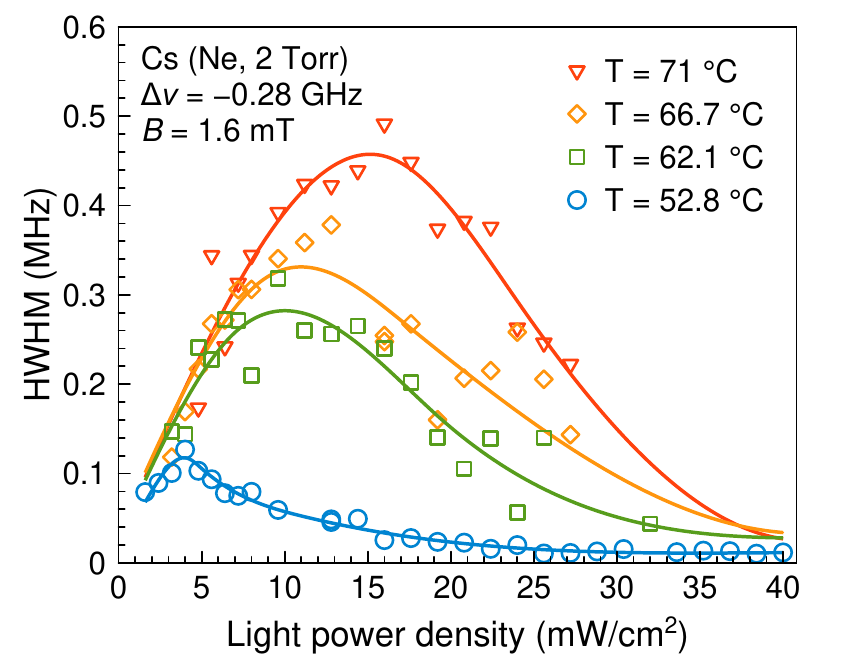}
	\caption{The width of the ``wide'' component of the SN resonance as a function of the probe light intensity at different temperatures: $T = 71$~\textdegree{}C (triangles), $T = 66.7$~\textdegree{}C (diamonds), $T = 62.1$~\textdegree{}C (squares), and $T = 52.8$~\textdegree{}C (circles). The solid lines are a guide to the eye.%
	}\label{fig8}
\end{figure}

Additional unexpected results were obtained from the measurements of temperature-related variations of the light-induced SN resonance broadening. Figure~\ref{fig8}  shows the light intensity dependence of the SN peak width at different temperatures of the cesium cell. 
For this figure, only the width of the ``wide'' component of the SN resonance was used. 
This made the dependence noisier but allowed us to separate the SN  contribution of the $F = 4$ state in a pure form and to make sure that admixture of spin fluctuations in the state $F = 3$ does not qualitatively affect the observed characteristics of the SN resonance broadening. 

Note that increasing the temperature mainly results in increased density of the Cs atoms in the cell, because changes in kinetic characteristics of atomic motion, in this temperature range, are evidently insignificant. 
Meanwhile, as we see from the figure, the increasing temperature generally strengthens the light-induced broadening of the SN resonance. 
This behavior of the SN linewidth under such low densities of cesium atoms, $(0.5 \ldots 2) \cdot 10^{12}$~cm$^{-3}$, also looks unusual and cannot be explained in a straightforward way. 

\section{Discussion}\label{sec:disc}

In this section, we discuss the key experimental results related to the probe-intensity variations of the SN resonance width and possible model interpretation of the observations.

\subsection{Model considerations}

Let us consider the effect of the light-induced broadening of the SN resonance in a more rigorous way. 
Note that, in the low probe-power regime, when the probe-induced perturbation of the spin noise may be neglected, the SN resonance width is determined by spontaneous relaxation processes in the system and, what is more important in our experimental conditions, by inhomogeneity of the applied magnetic field in the vapor cell~\cite{PhysRevA.97.032502, PhysRevResearch.2.012008, PhysRevA.103.023104}. 
The perturbations induced by the probe beam can be roughly separated into two classes: (i) the effects related to the renormalization of the energies of the spin states (mainly caused by the virtual transitions between the ground and excited spin multiplets and important for the detuned from the optical transition probe) and  (ii) the effects of real transitions of the atom between the ground and excited states (which are particularly prominent for the resonant probing). Generally, both effects coexist and should be taken into account simultaneously, but, for the sake of qualitative discussion, we address them separately. 

The effects of the renormalization of the spin states in the presence of the intense probe beam have been analyzed in our previous papers~\cite{Ryzhov:2016aa, PhysRevA.103.023104}. 
As shown in Ref.~\cite{PhysRevA.103.023104}, linearly polarized probe results in the light-induced splittings of the ground HF state $F=3$ and $F=4$. 
The magnitudes of the splittings depend on the probe intensity and orientation of the probe polarization plane with respect to the magnetic field. 
Additionally, if the probe beam is elliptically polarized, it produces the effective magnetic field controlled by the probe ellipticity~\cite{PhysRevLett.15.190, PhysRevA.63.043814, Ryzhov:2016aa}. 
Taking into account that the intensity of the probe beam can be inhomogeneous inside the cell (e.g., due to the Gaussian beam profile or due to absorption processes), the probe-induced energy renormalizations may result in an effective inhomogeneous broadening of the SN spectrum similar to the effect of the inhomogeneous magnetic field. 
While this kind of effects cannot be fully ruled out, we can note that, as was confirmed by our special measurements, the overall shifts of the SN peak induced by the probe were, in our experiments, smaller than the variation of the SN linewidth. 

Also, and importantly, these renormalizations can hardly explain the nonmonotonic dependence of the SN linewidth as a function of the probe beam intensity. 

Thus, it seems natural to relate the observed behavior of the SN resonance with the real transitions between the ground and excited states of the Cs atom induced by the intense probe beam. 
The role of the real transitions between the spin states can be illustrated in the framework of the simple phenomenological model that considers dynamics of the spin fluctuations in the ground, $\delta \bm F_g$, and excited, $\delta \bm F_e$, multiplets and the coupling between the multiplets via probe-induced optical transitions. We present the Bloch equations for the fluctuations in the form
\begin{subequations}
\label{fluct:bloch}
\begin{align}
\delta \dot{\bm F_g}+ \delta \bm F_g \times \bm \Omega_g +  \left(G+ \frac{1}{\tau_{s,g}}\right) \delta \bm F_g  - R \delta \bm F_e =\bm \xi_g,\\
\delta \dot{\bm F_e} + \delta \bm F_e \times \bm \Omega_e +  \left(R+ \frac{1}{\tau_{s,e}}\right) \delta \bm F_e  - G \delta \bm F_g =\bm \xi_e.
\end{align}
\end{subequations}
Here, the dot on top denotes the time-derivative, $\bm \Omega_g$ and $\bm \Omega_e$ are the spin precession frequencies in the ground and excited multiplets (these frequencies can, generally, include the contributions from the renormalization of the energy spectra), $\tau_{s,g}$ and $\tau_{s,e}$ are the spin-relaxation times unrelated to the optical processes,\footnote{The spin relaxation can be caused by interatomic collisions~\cite{PhysRevA.75.042502} or scattering on the cell walls. 
The times  $\tau_{s,g}$ and $\tau_{s,e}$ can phenomenologically include the contribution from the magnetic-field inhomogeneity, and, in the case of the excited state, the contributions to the transitions to another ground-state HF multiplet.} $G$ and $R$ are the optical excitation (pumping) and excited-state decay rates, respectively, and $\bm \xi_{g,e}$ are the random Langevin forces introduced in the theory of fluctuations~\cite{ll5_eng, ivchenko73fluct_eng, glazov2018electron}. 
In our simplest model, we assume that the spins of the ground and excited multiplets are the same ($F_g=F_e$) and optical processes are spin-conserving, with $G$ and $R$ being scalars,\footnote{The effects of spin relaxation and pumping in the course of optical transitions are disregarded, see, e.g.,~\cite{Aminoff1982, PhysRevB.89.081304} for details of these processes.} and set $R=G+1/\tau_0$, where $\tau_0$ is the spontaneous transition time. At a relatively weak probe intensity $I$, the optical pumping rate $G \sim \varpi_R^2 T_2/(1+\Delta^2T_2^2) \propto I$, where $\varpi_R$ is the Rabi frequency, $\Delta$ is the detuning between the probe beam frequency and the resonance frequency and $T_2$ is the optical coherence time. 
With the increase in $I$, the $G$ and $R$ saturate. 
Equations similar to the set~\eqref{fluct:bloch} have been used previously to describe the spin fluctuations in semiconductor quantum wells and quantum dots~\cite{PhysRevB.89.081304, glazov:sns:jetp16, PhysRevB.97.081403, PhysRevB.95.241408, glazov2018electron}, color centers~\cite{poshakinskii:YYY} and also the spin fluctuations and dynamics under electron hopping between the localization sites~\cite{PhysRevB.91.195301, PhysRevB.92.014206, PhysRevB.94.125305}. 
Note that the parameters in Eqs.~\eqref{fluct:bloch} should be considered as effective phenomenological parameters which depend, in particular, on the particular optical transition we study. 

\begin{figure}[t]
    \includegraphics[width=\linewidth]{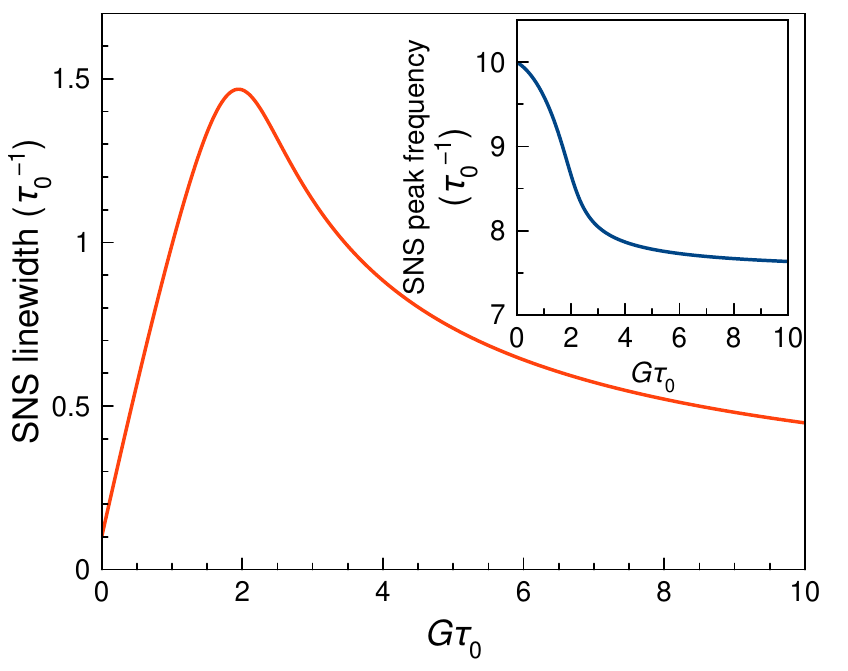}
    \caption{SNS linewidth (main panel) and peak position (inset) calculated after Eq.~\eqref{omega} as a function of the saturation parameter $G\tau_0$. The parameters of the calculations are $\Omega_g\tau_0 = 10$, $\omega_e\tau_0 = 5$, $\tau_0/\tau_{s,g}=0.1$, and $\tau_0/\tau_{s,e}=0.2$.%
    }\label{fig:theory}
\end{figure}

The complex eigenfrequencies $\omega_i = \omega_i' + \mathrm i \omega_i''$ of the equation set~\eqref{fluct:bloch} (at $\bm \xi_{g,e} \equiv 0$) describe the positions of the peaks, $\omega_i'$, and their widths, $ -\omega_i''$, in the SN spectrum. 
For the spin fluctuations transversal to the direction of the magnetic field, they can be found from the solution of the equation $\det{\mathcal K}=0$, where the matrix $\mathcal K$ describing  kinetics of the system reads
\begin{equation}
\label{K:mat}
\mathcal K =\begin{pmatrix}
-\mathrm i \omega + \mathrm i \Omega_g + G+ \tau_{s,g}^{-1} & - R\\
-G & -\mathrm i \omega + \mathrm i \Omega_e + R+ \tau_{s,e}^{-1} 
\end{pmatrix}.
\end{equation}
Such a form of $\mathcal K$ immediately follows from the two coupled equations for the $\delta F_g^+ =\delta F_{g,z} + \mathrm i \delta F_{g,y}$ and $\delta F_e^+ =\delta F_{e,z} + \mathrm i \delta F_{e,y}$, where we assumed that the magnetic field is applied along the $x$~axis, i.e., $\bm \Omega_e \parallel \Omega_g \parallel x$. 
It follows then that
\begin{multline}
\label{omega}
\omega_{\pm} = \frac{\Omega_g+\Omega_e - \mathrm i (\tau_{s,g}^{-1}+ \tau_{s,e}^{-1}) }{2} - \frac{\mathrm i }{2\tau_0} - \mathrm i G \\
\pm \mathrm i\frac{1}{2}\sqrt{\left[\frac{1}{\tau_0} + \frac{1}{\tau_{s,e}} - \frac{1}{\tau_{s,g}} + \mathrm i(\Omega_e - \Omega_g)\right]^2 +4 G\left(G+\frac{1}{\tau_0}\right)}.
\end{multline}

Let us analyze in detail the frequency $\omega_+$ which corresponds, at $G=0$, to the ground-state spin fluctuations. 
The dependence of the imaginary part of the frequency (i.e., the SN peak width) on $G$ for an arbitrary set of parameters is plotted in Fig.~\ref{fig:theory}. 
The inset shows the real part of the frequency (i.e., SN peak position)  as a function of the pumping rate $G$. 
At a weak probe, $G\tau_0 \ll 1$,
\begin{subequations}
\label{omega:limits}
\begin{multline}
\label{omega:weak}
\omega_+ \approx \Omega_g  - \frac{\mathrm i}{\tau_{s,g}}\\
 - \mathrm i  G\left(1 -\frac{1}{1+\tau_0/\tau_{s,e} - \tau_0/\tau_{s,g} + \mathrm i \tau_0(\Omega_e - \Omega_g)} \right),
\end{multline}
and both the spin precession frequency and the linewidth acquire linear in $G$ contributions resulting from the admixture of the excited-state dynamics. Depending on the relation between the system parameters, the SN resonance can be broader or narrower than at $G=0$. 
For the particular set chosen for Fig.~\ref{fig:theory}, the $|\Omega_e - \Omega_g|\tau_0 \gg 1$, and the SN resonance broadens $\propto G$. 
At the strong probe ($G\tau_0 \gg 1$), the dynamics of the fluctuations in the excited and ground state effectively averages, and
\begin{equation}
\label{omega:strong}
\omega_+ \approx \frac{\Omega_g+\Omega_e}{2} -  \frac{\mathrm i}{2\tau_{s,g}}- \frac{\mathrm i }{2\tau_{s,e}}.
\end{equation}
\end{subequations}
Thus, for sufficiently long $\tau_{s,e}$ the SN resonance width can decrease for sufficiently large pumping rates and, eventually, become comparable (or even smaller) than the SN linewidth measured at a weak probe.

This feature can be explained as follows. 
At sufficiently high values of the probe beam intensity, the processes of stimulated emission begin to prevail over the processes of spontaneous relaxation. 
As a result, the polarization properties of the photons absorbed and emitted by the atom become identical, and the atom returns to its initial state at the end of the optical excitation cycle. 
This effect is similar to the averaging of the spin precession frequencies due to the exchange interaction~\cite{PhysRevLett.89.130801, gis2014noise} or hopping~\cite{PhysRevB.91.195301, PhysRevB.92.014206}.

While the model consideration presented above allowed us, at least for a particular choice of the parameters, to obtain the dependence of the SN resonance width similar to that observed experimentally, the quantitative description of the data, including strongly pronounced spectral asymmetry of the effect across the optical transition, leave beyond the scope of the suggested simplified model. 
Still, the general pattern of the spectral and temperature behavior of the effect allows us to make certain assumptions about basic reasons underlying the observed anomalies. \looseness=-1
 
\subsection{Hypothetical assumption}
 
Experimental results of this work clearly show that spectra of optical perturbation of a spin system (as is often the case in nonlinear optics) may provide information about light-matter interactions hidden in linear optical spectra of the medium. 
As seen from  Fig.~\ref{fig4}, the optical transition $F = 4 \leftrightarrow F' = 3, 4, 5$, revealed in the linear SNS as a single homogeneously broadened spectral line~\cite{PhysRevA.97.032502}, under conditions of strong optical perturbation breaks down to two spectral regions with qualitatively different characteristics and a fairly sharp boundary between them. 
As seen from this figure, the short-wavelength region of the transition, characterized by the lower sensitivity of the SN resonance width to optical perturbation, occupies a smaller part of the line. 
We recall that the spectrum of the SN resonance broadening at the other HF component of the $\mathrm{D}_2$ line ($F = 3 \leftrightarrow F' = 2, 3, 4$) exhibits similar behavior with inverted short- and long-wavelength regions, Fig.~\ref{fig6}. 
This fact offers a clue to understanding such a structure of the SN perturbation spectrum. 
Referring to the energy diagram of the Cs atom, Fig.~\ref{fig1}, we can see that the short- and long-wavelength regions of the transitions from sublevels $F = 4$ and $F = 3$ ($F = 4 \leftrightarrow F' = 5$ and $F = 3 \leftrightarrow F' = 2$, respectively) correspond to the so-called ``closed'' transitions, when the excited atom can be radiatively deactivated only by returning back to the same state. 
When excited into other HF components of the upper state, the atom acquires the route of radiative relaxation to the other HF sublevel of the ground state. 
This process, which is known to be the reason for HF pumping~\cite{RevModPhys.44.169}, plays a crucial role, as we believe, in the effect of the light-induced SN resonance broadening. 
 
To ascribe to ``open'' transitions a higher sensitivity to optical perturbation of the ground-state spin-system, we have to accept that the process of optical excitation of the atom perturbs its ground-state spin system to a lesser degree when the atom returns back to its initial state. 
In other words, we have to accept that there exists a certain probability that the act of optical excitation does not change phase relations of the ground-state wave function of the atom. 
In the frame of the model presented above, it means, that, for the ``closed'' transitions, the $\tau_0$ and $\tau_{s,e}$ are significantly shorter than those of the ``open'' transitions, which makes the ground-state spin fluctuation dynamics much less sensitive to the excited-state dynamics for the case of the ``closed'' transitions. 
In this case, the escape from the excited optical transition (possible for the ``open'' transitions) will inevitably break the ground-state precession phase and will broaden the SN resonance. 
This is a strong assumption, and we did not manage to justify it theoretically, but our experimental results provide convincing evidence in favor of this hypothesis. \looseness=-1
 
The nonmonotonic behavior of the SN resonance broadening, as has been shown above, can be interpreted qualitatively with allowance for light-power-dependent dynamic characteristics of the excited state. 
Still, for a more accurate description of this dependence, we also have to take into account the effects of HF pumping and hole-burning {(disregarded in the model above)} that become essential at the elevated light-power densities used in our study. 
It is noteworthy that these two latter effects of bleaching of the cesium cell affect differently on the SN resonance broadening. 
With increasing the spectral hole depth, the number of resonance atoms decreases, the SN signal decreases also, but the main contribution to the SN signal is made by nonresonant and, therefore, unperturbed atoms, and the  SN resonance is getting narrower.  
The  HF pumping only decreases the SN signal amplitude. 

A specific feature of the temperature dependence of the effect, as shown in Fig.~\ref{fig8}, is the growth of the light-induced contribution to the SN resonance width with a temperature, i.e., with the density of cesium atoms. 
Since the laser field acting upon the atoms does not increase with increasing temperature, this behavior cannot be explained without taking into account interatomic interactions which may change either dynamic parameters of the excited state of cesium or the effective optical field due to the effects of radiation trapping. 
A complicated phenomenology of the discovered anomalies will also necessarily require taking into account a great variety of the effects of nonlinear magneto-optics in atoms \cite{RevModPhys.44.169, RevModPhys.74.1153}. 
We plan to investigate these effects in more detail elsewhere. 
 
\section{Conclusion}\label{sec:concl}
 
In this paper, we applied the method of nonlinear SNS for studying the sensitivity of the ground-state spin-system of cesium atoms to resonant optical perturbations. 
We have discovered and studied anomalous behavior of the SN resonance broadening of cesium under the condition of resonant probing and advanced a hypothesis about the effect of the excited-state relaxation route upon the efficiency of the spin-system optical perturbation. 
We have found, in particular, that optical perturbation of the spin-system through ``closed'' transitions proves to be less perturbative for the ground-state spin-system. 
We show also that the effects of the light-induced SN resonance broadening may be sensitive to the hidden HF structure of the excited state of the atom, as well as to interatomic interactions in the atomic medium. 
These findings, including narrowing of the magnetic resonance line under conditions of strong resonant probing, may be of interest both for atomic physics and for applications in magnetometry and metrology.

\vfill{}
 
\acknowledgements 
 
This work was supported by the RFBR Grant No. 19-52-12054 which is highly appreciated. 
The authors acknowledge Saint-Petersburg State University for a research grant 73031758.
M.M.G. acknowledges partial support of theoretical research by the RFBR Grant No. 19-52-12038. 

%

\bibliography{anomalous}

\end{document}